# Effective Ion Mobility and Long-Time Dark Current of Metal-Halide Perovskites of Different Crystallinity and Composition

Marisé García-Batlle[1], Sarah Deumel[2], Judith E. Huerdler[2], Sandro F. Tedde[2], Osbel Almora[1], and Germà Garcia-Belmonte[1]*

[1] *Institute of Advanced Materials (INAM), Universitat Jaume I, 12006 Castelló, Spain*

[2] *Siemens Healthineers AG, Technology Excellence, Guenther-Scharowsky-Strasse 1, 91058 Erlangen, Germany*

*\*Email: garciag@uji.es*

29 abril 2022

**Abstract**

Ion transport properties in metal-halide perovskite still constitute a subject of intense research because of the evident connection between mobile defects and device performance and operation degradation. In the specific case of X-ray detectors, dark current level and instability is regarded to be connected to the ion migration upon bias application. Different compositions ($MAPbBr_3$ and $MAPbI_3$) and structures (single- and micro-crystalline) are checked by the analysis of long-time dark current evolution. In all cases, electronic current increases with time before reaching a steady-state value within a response time (from $10^4$ s down to 10 s) that strongly depends on the applied bias. Our findings corroborate the existence of a coupling between electronic transport and ion kinetics that ultimately establishes the time scale of electronic current. Effective ion mobility $\mu_i$ is extracted that exhibits applied electrical field $\xi$ dependence that varies on the perovskite composition. While ion mobility results field-independent in the case of $MAPbI_3$, a clear field-enhancement is observed for $MAPbBr_3$ ($\partial \mu_i / \partial \xi > 0$), irrespective of the crystallinity. Both perovskite compounds present effective ion mobility in the range of $\mu_i \approx 10^{-7}$–$10^{-6}$ cm$^{-2}$ V$^{-1}$ s$^{-1}$, in accordance with previous analyses. The $\xi$-dependence of the ion mobility is related to the lower ionic concentration of the bromide compound.

Slower-migrating defect drift is suppressed in the case of MAPbBr$_3$, in opposition to that observed here for MAPbI$_3$.

# 1. Introduction

The enormous activity and research efforts in lead-halide perovskites materials has allowed to advance, not only in the field of solar cells and light emitters devices, but also in X-ray imaging and ionizing radiation detectors.[1, 2] Hybrid halide perovskite (HPs) materials have excellent optoelectronic properties, the structures can be tuned to adapt desired functionalities, the chemical formula $APbX_3$ (were A is a cation and X a halide anion) have reasonably high-Z elements which are helpful to stop high-energy radiation photons. HPs have the so-called defect tolerance which is connected to the transport properties allowing to maximize the product of the charge mobility and carrier lifetime by minimizing the bulk/surface defect density.[3-5] Furthermore, potentially direct semiconductor radiation detectors based on Pb-halide perovskites can be grown and processed in solution at relatively low temperature and from low-cost basics raw materials.[6, 7]

Fast and efficient detection of hard X- and γ-ray with high energy resolution is critical for medical and industrial applications.[8] The direct conversion X-ray detectors can transform the incident X-ray photons directly into electrical signals which can be an advantage in terms of high spatial resolution contrary to the indirect detectors (CsI).[9] Besides optimization in material composition and device architecture, the current conventional direct detectors, such as amorphous selenium (α-Se), suffer from their low mobility–lifetime ($\mu\tau$) product and small atomic number which limited their sensitivity. [5, 10] In contrast, HPs exhibited large $\mu\tau$-product and strong stopping power. However, there exist the persistent drawback of ion-migration which results in large noise with deleterious and instable dark current.[11-13] In fact, induced dark current and also photocurrent drift under a large electric field modify the intrinsic performance of the X-ray detectors.[14]

Many efforts have been made in order to develop a strategy in terms of atomistic surface passivation to heal the surface defects[15, 16], introducing charge transport layers at the outer interfaces[14], which may, potentially, inhibit the ion migration. However, high dark current level has been reported in perovskite-based detectors, still larger than that registered with commercial devices,[10, 17] over which the photocurrent should be detected. Moreover, registered dark current under continuous biasing exhibits instability, which is detrimental to the transient response of the X-ray detectors.[14] Ultimately, dark current levels are too large to achieve high quality images, with high resolution and contrast for accurate diagnosis.[18] Therefore, is essential to further progress in material composition engineering and device architecture to obtain high dark resistivity. Also crucial is achieving a deep understanding on how ion migration govern the long-time electronic current for the long-term operational stability of the detector.[19, 20]

In this work, chronoamperometry experiments are performed with the purpose of registering the long-time current transient response of different composition of single-crystal (SC) and micro-crystalline (MC) perovskite samples of $MAPbBr_3$ and $MAPbI_3$. Dissimilar current responses are encountered between the bromide-based perovskite samples and the iodine ones. Our findings corroborate the existence of a coupling between electronic transport and ion kinetics that ultimately establishes the time scale of electronic dark current.[21] In all cases, electronic current increases with time before reaching a steady-state value within a response time (from $10^4$ s down to 10 s) that strongly depends on the applied bias. We highlight that fitting the long-time dark current transient curve allows for the extraction of intrinsic parameters such as the ionic mobility which is in the range of $\mu_i \approx 10^{-7}$–$10^{-6}$ cm$^{-2}$ V$^{-1}$ s$^{-1}$ for the samples studied.

## 2. Results and Discussion

Long-time dark current transients of HPs samples of different thickness and crystallinity are shown in Figure 1. The transients were measured by following the protocol used previously,[21] by application of a forward-bias (for SC samples) and a reverse bias (for MC samples), followed by a zero-bias equilibration period (see Figure S1) of ~3000 s. In Figure S2, the general biasing protocol used for each sample is shown, which exhibit sufficient reproducibility. A general increasing trend can be seen (Figure 1), although some particularities were observed, at different times scales, depending on the sample composition and crystallinity.

By first examining the global transient response of a $MAPbBr_3$-SC (Figure 1a) under continuous application of forward bias, the current increases from a rather constant value at shorter times ($t < 1$–10 s), and finally saturates approaching steady-state values $J_1$ at longer times. This behaviors has been noticed previously in $MAPbBr_3$ single-crystal samples[21] and agrees with the evidence that as higher the bias is the faster the current increase. In Figure 1b, an unusual feature appears during the transient of a $MAPbBr_3$-MC sample under reverse bias. At the beginning of the transient for $t < 5$–50 s, a slight decrease of the current can be observed, while for $t > 100$ s an exponential growth, as in the SC, can be identified that finally reaches a current steady state $J_1$. As observed, the initial decrease represents a minor feature of the overall current evolution.

Figure 1c displays the complex current transient response of a MAPbI$_3$-MC sample. Once more, for shorter times ($t < 50$ s) initial current tend to slightly decrease. After that, current grows and seems to saturate at $J_1$ in a first exponential rise. This is consistent with previous observations in Figure 1 a and b registered for Br-based perovskite samples at that polarization times. Interesting, the transient response of a MAPbI$_3$-MC for bias lower than -10 V, is different than those observed at higher voltage steps: a second exponential rise appears, which tends to a larger saturation current $J_2$ at longer times.

Then, the magnitude of the electronic current transients $J$, both for the first and second exponential rise, can be described as a function of the time constant $\tau$ of the ion migration process as

$$J = J_i(1 - e^{-t/\tau_i}) \qquad (1)$$

being $J_i$ and $\tau_i$ ($i = 1, 2$) the corresponding saturation current and response time of each exponential rise step. Fitting of Equation 1 to the current rise is plot in each transient as a solid line in Figure 1. The time constants reveal the dynamics of the ionic transport within the device,[22] and can be assimilated to an drift-induced ionic time-of flight that effectively controls the time scale of the electronic current transient toward saturation, while the amplitude of the current response is determined by the electronic carriers (electrons and holes).[21]

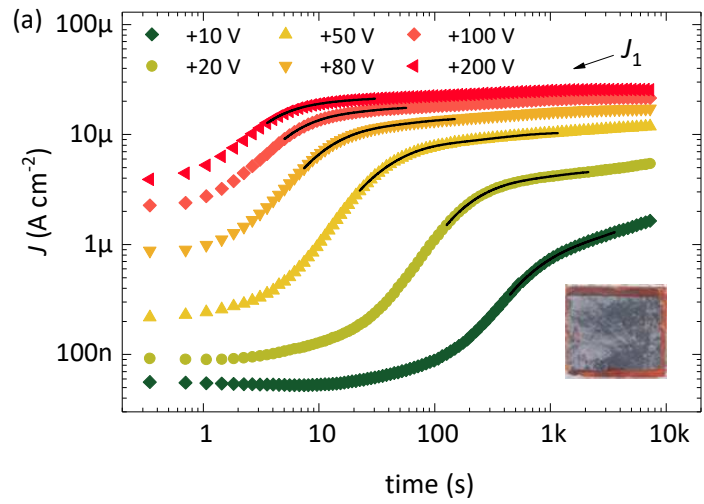

(a)

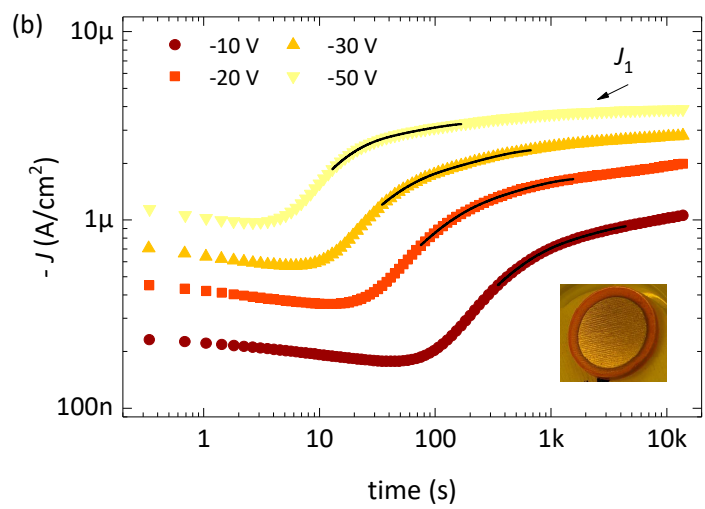

(b)

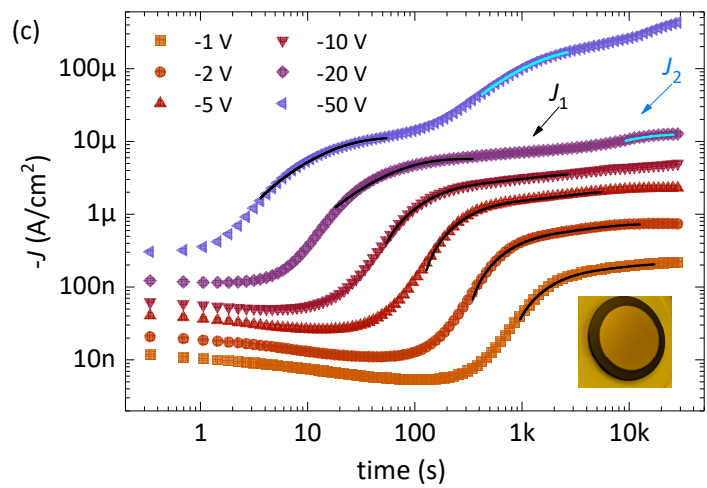

(c)

**Figure 1.** Long-time current response upon a biasing protocol for a) ~2 mm-thick MAPbBr$_3$-SC b) a ~1.5 mm-thick MAPbBr$_3$-MC c) ~1 mm-thick MAPbI$_3$-MC. Note the experimental current transient with the corresponding fittings (solid lines) for each curve following Equation 1. The parameters $J_1$ and $J_2$ correspond to the steady-state current during the first and second exponential rise respectively. After each bias, the device is kept under short-circuit (0-V bias) conditions to observe the relaxation current. See the long-time current response of each sample for the other two cycles in Figure S3, S4 and S5 respectively. In the inset is shown the top-view image of each sample. Data in a) reproduced with permission from ref.[21] (Copyright American Chemical Society).

Now let us consider the behavior of the steady-state electronic current. From the long-current transients of Figure 1, one can obtain the $J_1$–V curve in Figure 2 using the steady-state values of the first exponential rise. As noted, the current value $J_1$ has been marked in Figure 1 for the three samples presented. From an allometric fit with a power law of the type $J_1 \propto V^\beta$, the extracted parameter $\beta$ attains values between 0.98 and 1.05 with an average variability of ± 0.07, which suggests a seemly ohmic conductivity regime for electronic charge carriers within the explored bias range after current saturation. Note that steady-state current is plotted as a function of the applied electrical field $\xi = V/L$, being $L$ the sample thickness, for a better comparison among different structures. If the steady state values in Figure 2 are compared with the one obtained during a fast cyclic-voltammetry $(J - V$ curve) (Figure S6) one can deduced that steady-state values are higher than those encountered when faster scans are used. One example is observed in Figure S6b when the electronic current is instantly observed because of a much faster speed of electrons/holes compared to that of ions using a scan rate of 50 mV/s. That observation signals the necessary full ion relaxation before concluding about of electronic transport regimes in halide perovskite devices.[19]

We express the total characteristic time from the time constant $\tau$ of the current transient exponential growth as $t_t = 4\tau$, which relates to the 98% of the transition to the steady state (instead of the 63 % when only $\tau$ is used). For the sake of simplicity, one can assume a homogeneous ionic charge transport across the device. Under these conditions, the ionic time-of-flight mobility can be estimated through the simple form

$$\mu_i = \frac{L^2}{t_t V} \qquad (2)$$

In Figure 3a, for polycrystalline samples of different thickness of MAPbI$_3$, the ion mobility has been obtained by using the total time $t_t$ obtained first exponential rise (closed dots). Also plotted in Figure 3a is the $\mu_i$ estimated for the second exponential rise (open dots). It is remarked that ionic mobility exhibits a rather constant behavior with the increasing bias independently of the structure of the iodide samples.

On the contrary, Figure 3b shows $\mu_i$ for MAPbBr$_3$, with different crystallinity. One can infer different trend for bromide samples, compared to iodine ones: $\mu_i$ is observed to increase with the electric field $\xi = V/L$. From the analysis of Figure 3b, using an allometric fitting, one can calculate an exponent $m \sim 1.2 \pm 0.05$ of the relationship $\mu_i \propto \xi^m$. See below for more comments about this last point. We also notice that the use of Equation 2 assumes ion drift occurring along the whole sample thickness. If the actual ion transport takes place in narrow zones $L_{eff} < L$ (either internal bulk regions or in the vicinity of the external contacts related to the ionic Debye length), overestimated mobility values would be represented in Figure 3.

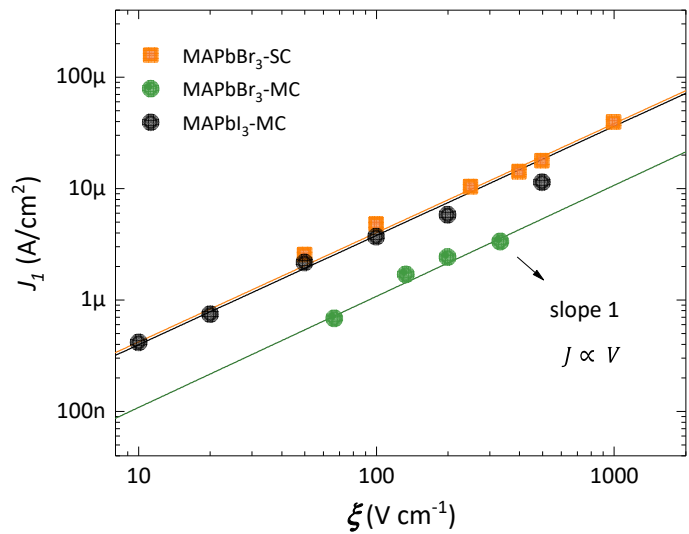

**Figure 2.** Steady state current during the first exponential rise for times $t > 100$ s of three different perovskite samples is shown with the corresponding linear fitting (solid lines). The ohmic character is remarkable ($J \propto V$)

On mixed conductors as HPs, one must be cautious about whether the measured signals are due to ionic or electronic defects, otherwise mistaken conclusions may be drawn.[23] Identifying the mobile defects is not easy and a number of reports predicts the migration of halide-related defects, under biasing conditions, rather than MA and Pb vacancies, with calculated activation barriers of ∼ 0.5 and 0.8 eV, respectively.[24, 25] Connections between specific ionic defects and electronic doping mechanisms have been

9made based on how the doping concentration is modified during the application and switching-off of an external bias.[21, 26] Concerning now Figure 1, one can speculate about being the halide vacancies $V_X^+$ the one that dominates and establish the general time response during the first exponential rise, with typical migration times in the range of 10–$10^4$ s. Only in the case of MAPbI$_3$-MC samples (Figure 1c), one can presume that the second exponential rise can be possibly originated by (*i*) imperfections on the perovskite surface or at the grain boundaries introducing alternative, and much slower, ion migration paths (*ii*) at higher electric fields, further enhancing the ion drift of slower-migrating defects, for instance $V_{MA}$ and $V_{Pb}$ or (*iii*) a combination of effects.

Even so, the electronic $J_1$ values in Figure 2 followed an ohmic regime irrespective of the sample crystallinity or composition, which may indicate that the electronic drift regime is proportional to the electrical field in most of the cases. Hence, our findings confirmed that the electronic-ionic coupling mechanism behind the current transients, rely in how the ionic movement establishes the kinetics of the electronic response in HPs materials.

Ionic mobility have been determined by many techniques such as: impedance spectroscopy (IS),[25-28] nuclear magnetic resonance (NMR) spectroscopy,[29, 30] photocurrent transient,[30-32] PL quenching method (PLQ),[33] temperature-dependent conductivity (TDC) measurement,[34, 35] and chronoamperometry measurements.[21, 27] Previous works shown values between $\mu_i \sim 1$–$3\times10^{-6}$ cm$^2$ V$^{-1}$ s$^{-1}$ for MAPbBr$_3$ perovskite single crystals by analyzing the resistance response during a diffusion-relaxation mechanism.[26] The same methodology has been applied for MAPbI$_3$ thick-pellets[27] obtaining self-consistent patterns by registering both, current transients and impedance spectra, with values encountered also in the range of ~ $10^{-6}$ cm$^2$ V$^{-1}$ s$^{-1}$.

Recalling now the resulting $\mu_i$ as a function of the electric field $\xi$ plotted in Figure 3a, one observes values within the range of $5\times10^{-7}$ to $3\times10^{-6}$ cm$^2$ V$^{-1}$ s$^{-1}$ for various micro-crystalline samples of MAPbI$_3$, which agree with those reported for the iodine-related defect ionic mobilities. Remarkably, a $\xi$-independent ion mobility is clearly found that would indicate a bulk origin, rather than interfacial, for the mechanism behind the observed electrical response. The $\mu_i$ calculated for the second exponential rise is found to be in the order of ~ $10^{-8}$ cm$^2$ V$^{-1}$ s$^{-1}$, with a similar, $\xi$-independent behavior. Therefore, it make sense to assume, for this second step, other possible moving ions responsible of the electronic response, such as MA$^+$ and Pb$^{2+}$ with much lower values of ion mobility.[36, 37] This is in accordance with the suppression of MA$^+$ migration in MAPbBr$_3$, while it is actually observed for MAPbI$_3$.[38] An alternative explanation for the second step in the transient response could be related to the role of grain boundaries. However, bromide-based devices, either of single- or micro-crystalline structure, do not exhibit such as feature at longer times, so as to move us to disregard it as structurally-originated.

By examining Figure 3b, a different trend occurs for bromide samples as $\mu_i \propto \xi^m$, compared to iodine ones. This field-dependent behavior deviates from the ohmic response expected when the ion velocity relates with the electrostatic potential as $v_d \propto \varphi$ that produces current patterns as $J \propto V$, and the time constants are directly proportional to the inverse of the applied bias $\tau \propto V^{-1}$.[21] Several models are suggested to rationalize for a field-dependent mobility behavior in different materials with $\partial \mu / \partial \xi > 0$. For instance, surface-charge decay in insulators were described to exhibit nonconstant carrier mobility.[39, 40] In the specific case of ion migration in electrolytes of different ionic strength, molecular dynamics simulations recently revealed that ionic conductivity is constant for strong electrolytes. On the contrary, weaker electrolytes or molten salts exhibit applied-field enhancement of the ion mobility.[41] Considering a perovskite sample at room temperature, the field-dependent mobility observed for MAPbBr$_3$ would indicate a weaker character for the bromide-related defect formation in opposition to a more favorable defect formation of iodide-related. Several theoretical predictions for MAPbBr$_3$ suggest a stronger Pb−Br bond[42] and considerable lattice contraction by stronger hydrogen bonding to the surrounding Pb−Br$_6$ octahedra.[38] These effects could, presumably, increase the bromide-related defect formation energy reducing the observed concentration of bromide mobile ions in MAPbBr$_3$ compared to iodide concentration in MAPbI$_3$,[38] which also explains the superior ambient stability of the bromide compound.

## 3. Conclusions

In summary, long-time dark current has been explored for a set of lead-halide perovskite thick samples of different crystallinity and composition. As a general trend, dark current exhibits an exponential-like rise that reaches steady-state values depending on the applied electrical field. Our findings reveal the coupling between slow ionic drift and the time-scale of the electronic current. MAPbBr$_3$ and MAPbI$_3$ exhibit different response: iodide compounds present double current rise related to the migration of much slower defects such as $MA^+$ and $Pb^{2+}$, with much lower values of ion mobility, that are instead suppressed in the case of $MAPbBr_3$. Also, the dependence of the ionic mobility on the electrical field presents dissimilar trends: constant, $\xi$-independent for $MAPbI_3$, but clearly $\xi$-dependent in the case of $MAPbBr_3$. Such a different behavior may be connected to difference in the chemistry of the defect formation between both compounds.

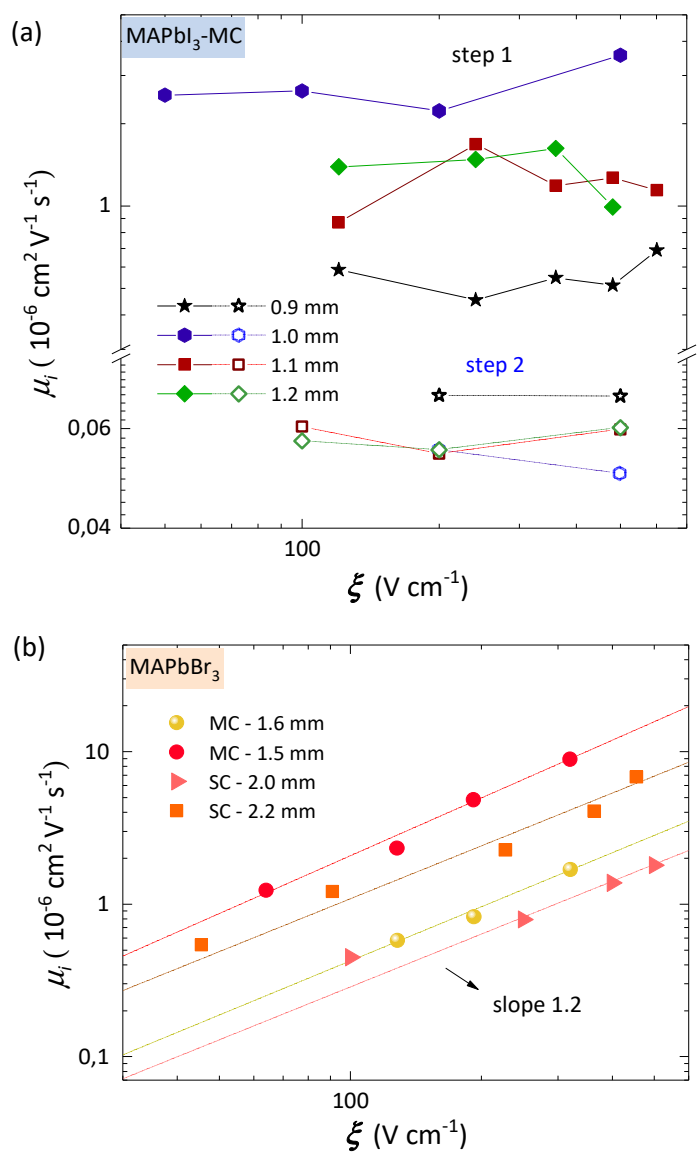

**Figure 3.** Ion mobility as a function of the electric field for a) various $MAPbI_3$-MC perovskite samples and b) two $MAPbBr_3$-SC and two MC perovskite samples. While a linear trend is clearly observed for bromide compounds, no correlation appears for iodide perovskites.

## 4. Experimental section

*Sample preparation:* Perovskite samples of two compositions, MAPbBr$_3$ and MAPbI$_3$, were analyzed. Firstly, single crystals (SC) of MAPbBr$_3$ symmetrically contacted with Cr electrodes (see Figure 1a, inset) were prepared following the inverse temperature crystallization (ITC) growth method previously reported.[43] Secondly, MAPbX$_3$ (X =Br, I) micro-crystalline pellets (MC) of ~1 mm-thick asymmetrically contacted with evaporated Pt and Cr electrodes of 1 cm$^2$ area (see Figure 1b and c, inset) were made by a soft-sintering process, which is described in detail in ref.[44] were prepared. Table S1 in the Supplemental Information summarizes the general characteristics of the samples studied. More information about the structural characterization and optoelectronics properties can be found in references [26, 43, 45] for MAPbBr$_3$-SC and [27, 44] for MAPbI$_3$-MC samples.

*Electrical Measurements:* Chronoamperometry measurements were carried out with a Source Measure Unit Model 2612B from Keithley Instruments, Inc. Current measurements were conducted following the long-time direct-current mode (DC) bias protocol in the ranges of ± 200 to 0 V as previously reported.[21] The samples were kept in the dark at ambient temperature and with N$_2$ circulation to avoid humidity- and oxygen-induced degradations.

Data available on request from the authors.

**Supporting Information.**

**Acknowledgments**

This work has received funding from the European Union's Horizon 2020 research and innovation program under the Photonics Public Private Partnership (www.photonics21.org) with the project PEROXIS under the grant agreement N° 871336. Funding for open access charge: CRUE-Universitat Jaume I.

# *Supporting Information*

# Effective Ion Mobility and Long-Time Dark Current of Metal-Halide Perovskites of Different Crystallinity and Composition

Marisé García-Batlle[1], Sarah Deumel[2], Judith E. Huerdler[2], Sandro F. Tedde[2], Osbel Almora[1] and Germà Garcia-Belmonte[1]*

[1] *Institute of Advanced Materials (INAM), Universitat Jaume I, 12006 Castelló, Spain*

[2] *Siemens Healthineers AG, Technology Excellence, Guenther-Scharowsky-Strasse 1, 91058 Erlangen, Germany*

*Email: garciag@uji.es

29 April 2022

| Table S1. Characteristics of the samples studied | | | | | | |
|---|---|---|---|---|---|---|
| **Sample Composition** | **Number of samples** | **Cristallinity** | **Dimensions** | **Sample thickness (μm)** | **Electrode configuration and area** | **Picture of the sample** |
| MAPbI$_3$ | 6 | MC | diameter ~15 mm | ~1000 | Pt/Cr ~ 1cm$^2$ | 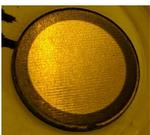 |
| MAPbBr$_3$ | 2 | MC | diameter ~15 mm | ~1560 | Pt/Cr ~ 1cm$^2$ | 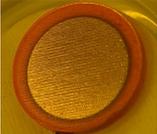 |
| MAPbBr$_3$ | 2 | SC | 3.93 mm × 3.87 mm | ~2000 | Cr/Cr ~ 0.12 cm$^2$ | 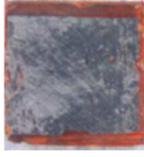 |

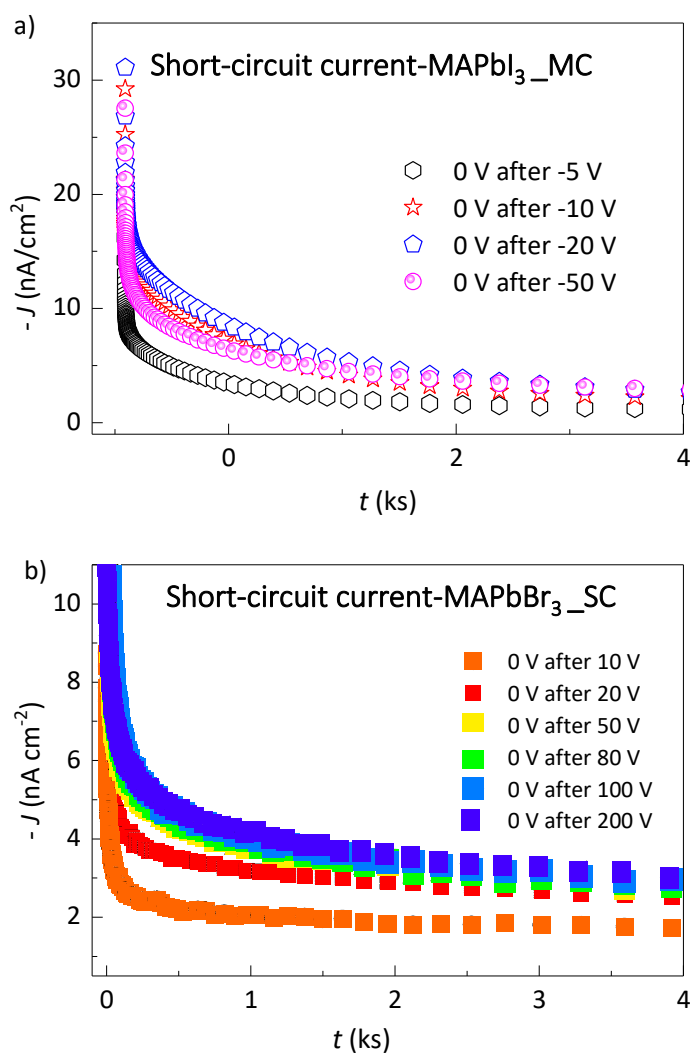

**Figure S1** Example on current transient response to short-circuit condition (0 V-bias voltage) of a a) MAPbI$_3$ MC sample contacted with Pt/Cr electrodes b) a MAPbBr$_3$ SC symmetrically contacted with evaporated Cr electrodes. Short circuit current

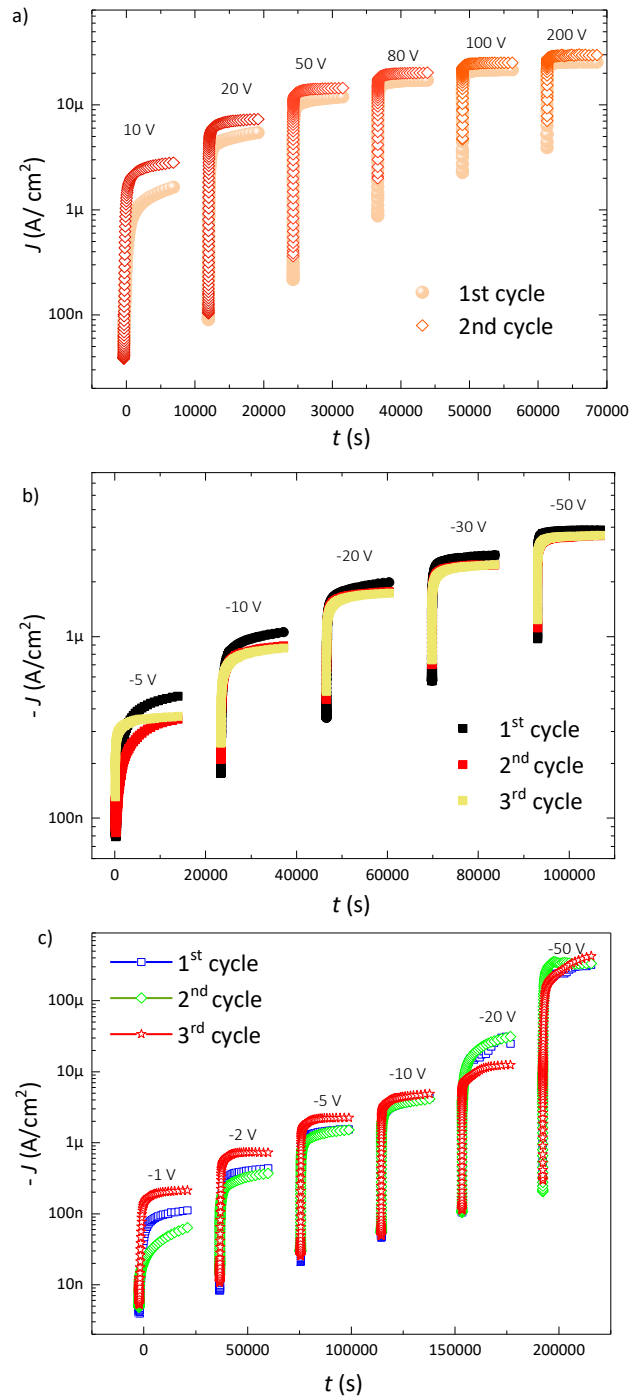

**Figure S2.**) Protocol of measurement based on long-time current transient's response to different voltage steps of a) a 2.2 mm MAPbBr$_3$ SC- symmetrically contacted with evaporated Cr electrodes and two samples of b) MAPbBr$_3$ MC and c) MAPbI$_3$ MC contacted with Pt/Cr electrodes. Note the reproducibity bewteen cycles.

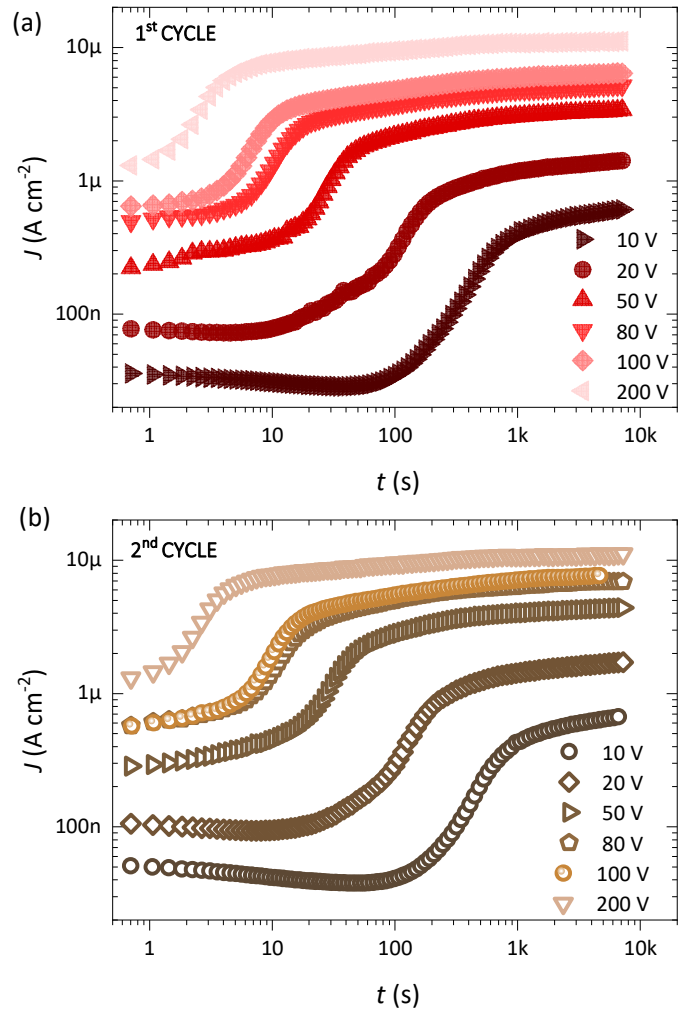

**Figure S3** ) Long-time current transient response to different voltage steps during the 1st and b) 2nd cycle of measurement of a 2 mm-thick MAPbBr$_3$ SC.

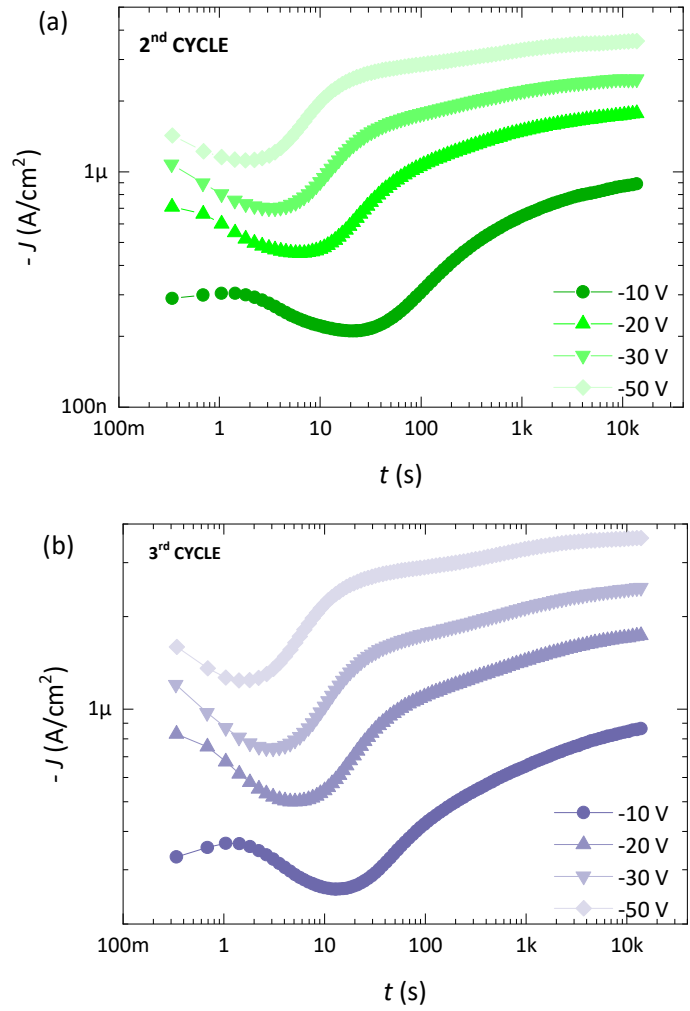

**Figure S4**. Long-time current transient response to different voltage steps during the 2nd and b) 3rd cycle of measurement of a 1.5 mm-thick MAPbBr$_3$ MC.

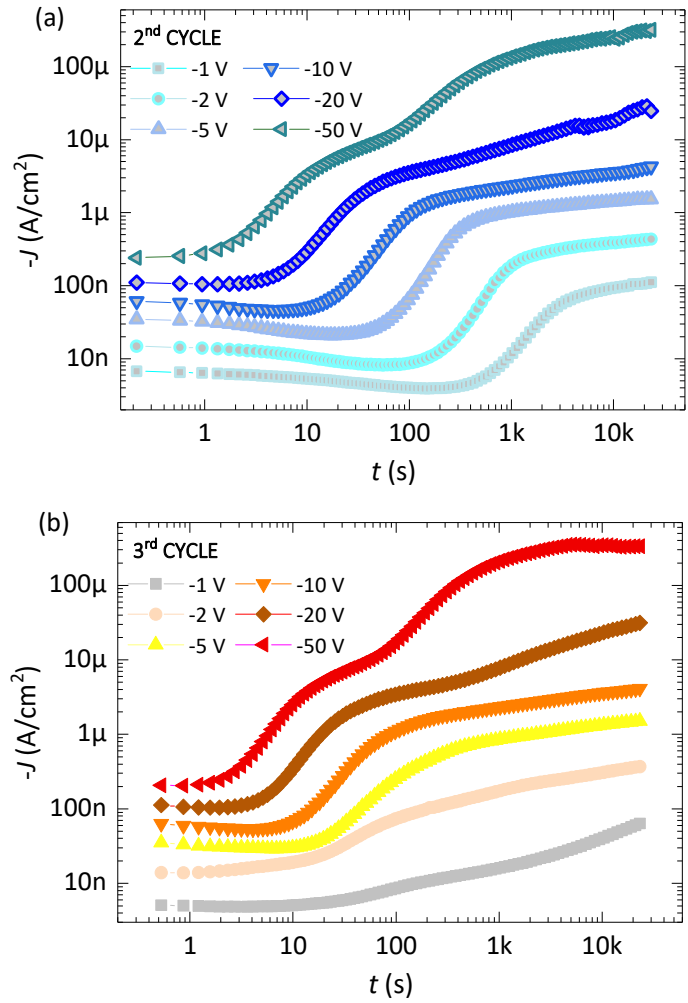

**Figure S5**. Long-time current transient response to different voltage steps during the 2nd and b) 3rd cycle of measurement of a 1 mm-thick MAPbI$_3$ MC.

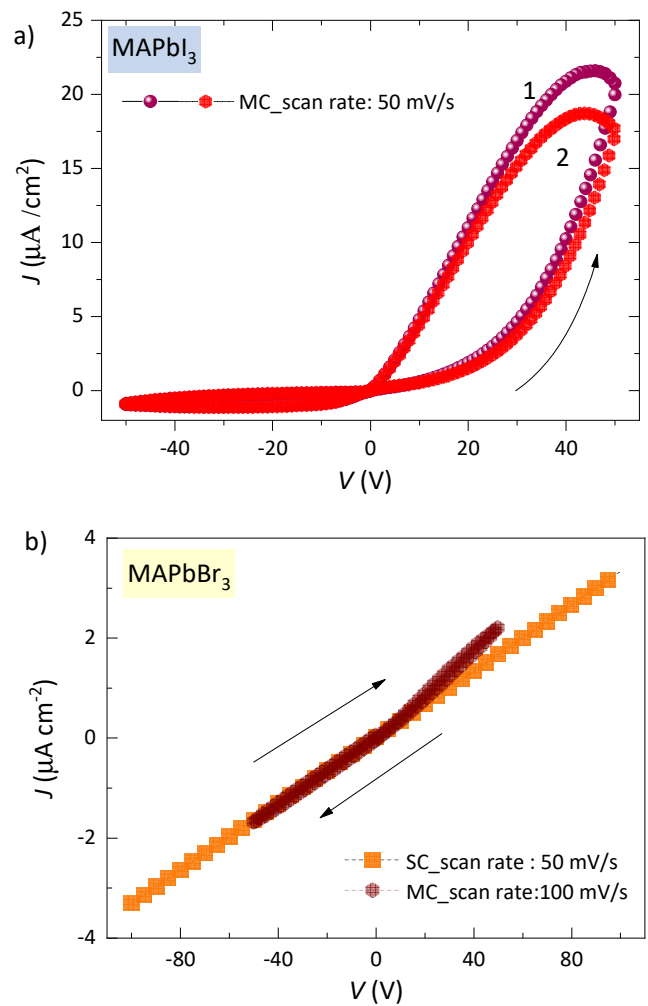

**Figure S6.** Current–voltage characteristics ($j$-$V$) with scan rate of a) 50 mV/s and step of 1 V of 1 mm-thick MAPbI$_3$ MC sample b) 50 mV/s and 100 mV/s and step of 1 V of firstly, a 2 mm-thick MAPbBr$_3$ SC and secondly a 1.5 mm-thick MAPbBr$_3$ MC. In Fig b it is remarkable the ohmic character of the characteristics $j$-$V$ curve, in agreement with previous analysis on Cr-contacted perovskite device[1]